\documentclass[twocolumn,pre]{revtex4}
\usepackage{amssymb,amsmath,bm}
\usepackage[dvips]{graphicx}
\usepackage{setspace}
\usepackage[dvips]{color}

\begin{document}

\title{Nonlinear dielectric effect of dipolar fluids}

\date{\today}

\author{I. Szalai}
\email{szalai@almos.vein.hu}
\affiliation{Institute of Physics, University of Pannonia, 
H-8201 Veszpr\'em, POBox 158, Hungary}

\author{S. Nagy}
\affiliation{Institute of Physics, University of Pannonia, 
H-8201 Veszpr\'em, POBox 158, Hungary} 

\author{S. Dietrich}
\email{dietrich@mf.mpg.de}
\affiliation{Max-Planck-Institut f\"ur Metallforschung, Heisenbergstr. 3, D-70569 Stuttgart, Germany}
\affiliation{Institut f\"ur Theoretische und Angewandte Physik, Universit\"at Stuttgart, Pfaffenwaldring 57, D-70569 Stuttgart, Germany}

\begin{abstract}
The nonlinear dielectric effect for dipolar fluids is studied within the framework of the mean spherical approximation (MSA) of hard core dipolar Yukawa fluids. Based on earlier results for the electric field dependence of the polarization our analytical results show so-called normal saturation effects which are in good agreement with corresponding NVT ensemble Monte Carlo simulation data. The linear and the nonlinear dielectric permittivities obtained from MC simulations are determined from the fluctuations of the total dipole moment of the system in the absence of an applied electric field. We compare the MSA based theoretical results with the corresponding Langevin and Debye-Weiss behaviors.
\end{abstract}

\maketitle

\section{Introduction}
The studies of dielectric polarization in fluids \cite{bo1} are based on the following relation between the polarization $\mathbf{P}$ and the electric field strength $\mathbf{E}$ inside the dielectric: 
\begin{equation}
4\pi\mathbf{P}=(\epsilon_E-1)\mathbf{E},
\label{fpol}
\end{equation}
where $\epsilon_E$ is the field dependent dielectric permittivity. The internal field is often called Maxwell field \cite{bo1,di1} which differs from the external field $\mathbf{E}_0$ applied to the dielectric medium. In the low-field limit the linear dielectric permittivity $\epsilon_0$ of an isotropic fluid is given by the ratio of the polarization to the internal field strength:
\begin{equation}
\epsilon_0=1+4\pi\lim_{E\rightarrow{0}}\frac{P}{E}
=1+4\pi\left(\frac{\partial{P}}{\partial{E}}\right)_{E=0}.
\label{ep0}
\end{equation}
In strong electric fields the polarization response acquires in addition nonlinear contributions, so that the electric permittivity turns into a nonlinear function of the Maxwell field: 
\begin{equation}
\epsilon_E=\epsilon_0 +\epsilon_2{E}^2+\epsilon_4{E}^4+...\,\,\,\,\,\,\,.
\label{sor}
\end{equation}
For simple molecular liquids the coefficients of the power series  decrease rapidly \cite{rz1}. Thus in general dielectric experiments can be well described in terms of the linear dielectric permittivity. For liquids consisting of small molecules this series 
can be limited to the second term. The corresponding nonlinear dielectric effect (NDE) is determined by the coefficient $\epsilon_2$ of the contribution  $\sim{E^2}$ in Eq. (\ref{sor}):
\begin{equation}
\epsilon_2\equiv 4\pi\lambda=
\lim_{E\rightarrow0}\frac{\epsilon_E-\epsilon_0}{E^2}=
\lim_{E\rightarrow0}\frac{\Delta\epsilon_E}{E^2},
\label{lam}
\end{equation}
where $\lambda$ is the nonlinear dielectric permittivity which, upon substituting Eq. (\ref{sor}) into Eq. (\ref{fpol}), is given by
\begin{equation}
\lambda=\frac{1}{3!}\left(\frac{\partial^3{P}}{\partial{E^3}}\right)_{E=0}.
\label{der3}
\end{equation}

Although the nonlinear dielectric behavior of liquids has enjoyed a long lasting scientific interest \cite{he1,pi1} accurate measurements of NDEs have become possible only recently due to the development of new techniques \cite{rz1}, which in an applied electric field are capable to separate from the NDE, e.g., the Joule effect due to heating of the sample induced by its finite conductivity.
Experimentally, $\lambda$ is determined from the small change of the dielectric permittivity ($\Delta\epsilon_E$) induced by static \cite{he1,pi1} or pulsed \cite{ma1,sr1,rz1,rz2} strong electric fields and is detected by a weak radio-frequency probing field. In nonlinear optics the NDE can also provide useful information for laser induced molecular reorientations in isotropic and liquid crystalline phases \cite{kh1}.
The first NDE measurements were carried out by Herweg in diethyl ether \cite{he1} yielding $\epsilon_2<0$. If $\epsilon_2$ has a sign opposite to that of $\epsilon_0>0$ one often speaks of normal saturation because in this case the first correction term to the linear behavior $P{\sim}E$ for $E{\rightarrow}0$ is in line with the levelling off at large $E$.
This effect is mainly connected to the ordering of the orientation of dipoles in strong electric fields. The negative value of $\epsilon_2$ corresponds to the negative value of the third order term of the power expansion of the Langevin function for small $E$. 
Strongly dipolar liquids, such as nitrobenzene, show anomalous (positive) dielectric saturation because in such systems the external electric field influences the formation of
antiparallel pairs of dipolar nitrobenzene molecules \cite{pi1}. Nonlinear dielectric effects have become also very useful for analyzing intermolecular association \cite{ma2}, conformational equilibria \cite{no1}, and critical phenomena in liquids and liquid mixtures \cite{rz3,rz4}.
Recently the NDE has also been used to study isotropic - mesophase transitions of 
various liquid crystals \cite{dr1}.

The theoretical models of nonlinear dielectric phenomena are based on classical electrostatics and statistical mechanics of liquids.
There have been early attempts by Debye \cite{de1}, Onsager \cite{on1}, and Kirkwood \cite{ki1} to calculate the nonlinear dielectric permittivity on the basis of phenomenological theories of dielectric continua. Nice summaries of these nonlinear theories can be found in Refs. \cite{bo1} and \cite{co1}.
From a microscopic point of view Rasaiah et al. \cite{ra1} and Martina and Stell \cite{st1} have proposed a statistical mechanics description for NDE and 
electrostriction on the basis of a quadratic hypernetted chain approximation.
Using molecular dynamics (MD) simulations the dielectric saturation of water has been studied by Alper and Levy \cite{al1}, but they have not published any numerical value for the nonlinear dielectric permittivity. Yeh and Berkowitz \cite{ye1} have obtained new MD simulation data for the electric field dependence of the dielectric permittivity of water and they found that $\epsilon_E$ decreases with increasing applied field strength in accordance with the expectation of a normal saturation effect. Their external field simulation data show good agreement with a phenomenological equation proposed by Booth \cite{bh1}. This equation has been used in the calculation of dielectric saturation of water in membrane protein channels \cite{ag1}.
Recently Fulton \cite{fu1} has compared the nonlinear dielectric permittivity of water obtained from simulation, theory, and experiment. He concluded that an upgraded approach by Booth \cite{bh2} renders the best agreement between the simulation and experimental data; but the extent of the agreement depends on the water model used for determining the corresponding correlation functions. It is also shown that the calculated nonlinear dielectric permittivity strongly depends on the application of phenomenological cavity and reaction field corrections. These findings underscore the ongoing interest in statistical mechanics analyses of nonlinear dielectric effects of dipolar liquids.

Within the framework of density functional theory (DFT) and the mean spherical approximation (MSA) we have proposed an equation \cite{sz1} for the magnetic field dependence of the magnetization of ferrofluids, which turned out to be successful in comparison with corresponding Monte Carlo (MC) simulation data. Translated into the
synonymous electric language this means that we have obtained an analytical (implicit) equation for the electric field dependence of the polarization. 
Here, from our previous results \cite{sz1}, we deduce a formula for the nonlinear dielectric permittivity of dipolar fluids. Moreover, we compare our theoretical findings with canonical MC simulation data using the corresponding fluctuation formulae in the absence of external fields. The motivation for our study is to deduce an analytical theory of a realistic model for nonlinear phenomena in dipolar liquids. This theory has to be quantitatively correct in the sense that it withstands the comparison with MC simulation data and it has to be easily applicable for interpreting corresponding experimental data. We expect that our model calculations can be used to shed light on the role of nonlinear dielectric saturation for solvation effects of ions in dipolar solvents \cite{ou1,ou2}.

\section{Theory}
\subsection{Microscopic model}
We study hard core dipolar Yukawa fluids which consist of spherical particles interacting via a hard core Yukawa (Y) potential characterized by parameters $\sigma$, $\varepsilon_Y$, and $\kappa$:

\begin{equation}
u_{Y}(r_{12})=\left \{
      \begin{array}{lll}
	 \infty &, &  r_{12}<\sigma \\
         -\varepsilon_Y\sigma{(r_{12})^{-1}}\exp[-\kappa(r_{12}-\sigma)] &, & 
         r_{12}\geq\sigma .\\
       \end{array} 
            \right.
\label{usw}
\end{equation}
In addition there is a dipolar interaction due to point dipoles embedded at the centers of the particles:
\begin{equation}
u_{D}(\mathbf{r}_{12},\omega_1,\omega_2)=-\frac{m^2}{r_{12}^3}D(\omega_{12},\omega_1,\omega_2),
\end{equation}
with the rotationally invariant function 
\begin{equation}
D(\omega_{12},\omega_1,\omega_2)=
3(\mathbf{\widehat{m}}_1\cdot\mathbf{\widehat{r}}_{12})
(\mathbf{\widehat{m}}_2\cdot\mathbf{\widehat{r}}_{12})
-(\mathbf{\widehat{m}}_1\cdot\mathbf{\widehat{m}}_2),
\label{DDD}
\end{equation}
where particle 1 (2) is located at $\mathbf{r}_1$ ($\mathbf{r}_2$) and carries a dipole moment of strength $m$ with an orientation given by the unit vector $\mathbf{\widehat{m}}_1(\omega_1)$ ($\mathbf{\widehat{m}}_2(\omega_2)$) with polar angles $\omega_1=(\theta_1,\phi_1)$ ($\omega_2=(\theta_2,\phi_2)$); $\mathbf{r}_{12}=\mathbf{r}_1-\mathbf{r}_2$ is the difference vector between the centers of particle 1 and 2, $r_{12}=|\mathbf{r}_{12}|$, and $\mathbf{\widehat{r}}_{12}=\mathbf{r}_{12}/r_{12}$ is a unit vector with orientation $\omega_{12}=(\theta_{12},\phi_{12})$.
The hard core dipolar Yukawa interaction potential is defined by the sum of the aforementioned potentials as
\begin{equation}
u_{DY}(\mathbf{r}_{12},\omega_1,\omega_2)=u_Y(r_{12})+u_D(\mathbf{r}_{12},\omega_1,\omega_2).
\end{equation}
\subsection{MSA for dipolar Yukawa fluids}
The MSA is defined by three equations relating the total correlation function $h(\mathbf{r}_{12},\omega_1,\omega_2)$ and the direct correlation function $c(\mathbf{r}_{12},\omega_1,\omega_2)$ as follows:
\begin{eqnarray}
h(\mathbf{r}_{12},\omega_1,\omega_2)=c(\mathbf{r}_{12},\omega_1,\omega_2)+\nonumber\\
\frac{\rho}{4\pi}\int{d\omega_3}\int{d^3r_3}
h(\mathbf{r}_{13},\omega_1,\omega_3)c(\mathbf{r}_{23},\omega_2,\omega_3),
\label{oz}
\end{eqnarray}
\begin{equation}
h(\mathbf{r}_{12},\omega_1,\omega_2)=-1,\,\,\,\,\,r_{12}<\sigma,
\label{oz1}
\end{equation}
and
\begin{equation}
c(\mathbf{r}_{12},\omega_1,\omega_2)=-\beta{u(\mathbf{r}_{12},\omega_1,\omega_2)},
\,\,\,\,\,r_{12}\geq\sigma.
\label{oz2}
\end{equation}
Here and in the following $\beta=1/k_BT$, where $k_B$ is the Boltzmann constant and $T$ is the temperature, and $u(\mathbf{r}_{12},\omega_1,\omega_2)$ is the pair potential characterizing the system. The number density $\rho=N/V$ is given by the number $N$ of molecules in the system of volume $V$.
Equation (\ref{oz}) is the Ornstein-Zernike (OZ) relation, Eq. (\ref{oz1}) is an exact relation for hard spheres, and Eq. (\ref{oz2}) is the closure relation for the MSA (for details see Ref. \cite{ha1}).
For dipolar hard sphere (DHS) fluids an analytical solution of the MSA was reported by Wertheim \cite{we1}. Later the MSA for hard core Yukawa fluids has also been solved analytically by Waisman \cite{wa1}. Following the ideas of Wertheim, for hard core dipolar Yukawa fluids Henderson et al. \cite{sz2,sz3} found an analytical solution in the framework of MSA. Within this theory the direct correlation function of the DY fluid can be expressed as
\begin{eqnarray}
c_{DY}(\mathbf{r}_{12},\omega_1,\omega_2)=c_Y(r_{12})+\nonumber\\
c_D(r_{12})D(\omega_{12},\omega_1,\omega_2)
+c_{\Delta}(r_{12})\Delta(\omega_1,\omega_2),
\label{cdy}
\end{eqnarray}
where
\begin{equation}
\Delta(\omega_1,\omega_2)=\mathbf{\widehat{m}}_1\cdot\mathbf{\widehat{m}}_2
\label{Del} 
\end{equation}
is a rotationally invariant function. (We note that a similar equation is valid for the
total correlation function $h_{DY}(\mathbf{r}_{12},\omega_1,\omega_2)$ of the DY fluid.) 
The radially symmetric function $c_Y(r_{12})$ is the solution of the OZ equation with a simple hard-core Yukawa MSA closure and is given in Refs. \cite{wa1,hh1} while the functions $c_D(r_{12})$ and $c_{\Delta}(r_{12})$ are the solutions of two MSA integral equations (derived from the OZ equations) with the  corresponding dipolar hard sphere MSA closure, depending therefore on the dipole moment of the molecules \cite{we1}. In Eq. (\ref{cdy}) $D$ and $\Delta$ are given 
by Eqs. (\ref{DDD}) and (\ref{Del}) respectively. The coefficients $c_D(r_{12})$ and $c_{\Delta}(r_{12})$ (together with $h_{D}(r_{12})$ and $h_{\Delta}(r_{12})$ ) are independent of the Yukawa potential parameters $\varepsilon_Y$ and $\kappa$ as well as of $\omega_{12}$, $\omega_1$, $\omega_2$ and are given in Ref. \cite{we1}. The main feature of this solution is that it decomposes into contributions from the Yukawa potential and from the dipolar hard sphere potential with the latter ones factorizing into radial and angular dependences. 
In Ref. \cite{sz2} it has been shown that, within the framework of MSA, the free energy $F_{DY}$ of the DY fluids can be written as
\begin{equation}
F_{DY}=F_{ID}+F_{DY}^{ex}=F_{ID}+F_{HS}^{ex}+F_{Y}^{ex}+F_{DHS}^{ex},
\label{fdyuk}
\end{equation}
where $F_{ID}$ is the ideal gas free energy and $F_{HS}^{ex}$, $F_{Y}^{ex}$, and $F_{DHS}^{ex}$ are the excess free energies of hard sphere \cite{ha1}, hard core Yukawa \cite{wa1}, and dipolar hard sphere fluids \cite{we1}, respectively. Within this approximation the dielectric constant of the DY fluid is given by the formula due to Wertheim \cite{we1} and Henderson et al. \cite{sz2}: 
\begin{equation}
\epsilon_0=\frac{q(2\xi(y))}{q(-\xi(y))},
\label{ep0msa}
\end{equation}
where
\begin{equation}
q(x)=\frac{(1+2x)^2}{(1-x)^4}
\end{equation}
is the reduced inverse compressibility function of the hard sphere fluid. The parameter $\xi(y)$ stems from the DHS MSA and is given by the implicit equation
\begin{equation}
3y=q(2\xi)-q(-\xi),
\label{yp}
\end{equation}
where 
\begin{equation}
y=\frac{4\pi}{9}\frac{{m^2}\rho}{k_BT}
\label{yyy}
\end{equation}
measures the reduced dipole strength.
\subsection{Field dependence of polarization}

The disadvantage of the MSA is that it can predict the polarization only for weak electric fields because in essence it is a linear response theory. 
In order to overcome this shortcoming we resort to density functional theory. 
Within this framework we first have to determine the orientational distribution function in an applied external field. For an inhomogeneous and anisotropic dipolar fluid the one-particle distribution function ${\rho}(\mathbf{r},\omega)$ depends on the position $\mathbf{r}$ and the orientation $\omega$ of the particles. Accordingly the number density is given by ${\rho}({\bf{r}})=\int{d\omega}{\rho}(\mathbf{r},\omega)$ 
and $\alpha(\mathbf{r},\omega)={\rho}(\mathbf{r},\omega)/\rho(\bf{r})$ is the orientational distribution function. In a homogeneously polarized bulk phase (concerning influences from the sample shape see below) the number density $\rho$ is spatially constant and the orientational distribution function $\alpha({\bf{r}},\omega)=\alpha(\theta)$ depends only on the angle $\theta$, which measures the orientation of the dipole of a single particle relative to the field direction. (Here we do not consider electrostriction, i.e., the dependence of $\rho$ on $E$, because this contribution to $\Delta\epsilon$ is one order of magnitude smaller than the corresponding contribution of the orientational ordering of dipoles in the presence of an applied electric field \cite{kr1}).
The orientational distribution function can be obtained by minimizing the following grand canonical variational functional:
\begin{eqnarray}
\Omega[\rho,\{\alpha(\theta)\},T,\mu]=F_{ID}[\rho,\{\alpha(\theta)\},T]+\nonumber\\
F^{ex}_{DY}[\rho,\{\alpha(\theta)\},T]-\nonumber\\
\rho\int{d^3r}{d\omega}\alpha(\theta)(\mu+mE\cos{\theta}),
\label{gra}
\end{eqnarray}
where $F_{ID}$ and $F_{DY}^{ex}$ are the ideal gas and the excess dipolar Yukawa free energy density functionals, respectively, $\mu$ is the chemical potential, and $E$ is the internal electric field in the sample. In this theory we assume that the volume of the fluid $V$ has the shape of a macroscopic prolate rotational ellipsoid (elongated around the electric field direction) so that the internal field strength $E$ is equal to the strength of the externally applied electric field $E_0$ (see Ref. \cite{di1}).
The excess DY free energy functional for an anisotropic system is not known. However, it can be approximated by a functional Taylor series, expanded around a homogeneous isotropic reference system with bulk number density $\rho$. Neglecting all terms beyond second order, one has
\begin{eqnarray}
{\beta}F^{ex}_{DY}[\rho,\{\alpha(\theta)\},T]={\beta}F^{ex}_{DY}(\rho,T)-\nonumber\\
\frac{\rho^2}{2}\int{d^3r_1}{d\omega_1}\int{d^3r_2}{d\omega_2}
\Delta\alpha(\theta_1)\Delta\alpha(\theta_2)\times\nonumber\\
c_{DY}(\mathbf{r}_{12},\omega_1,\omega_2),
\label{ser}
\end{eqnarray}
where $\Delta\alpha(\theta)=\alpha(\theta)-1/(4\pi)$ is the difference between the actual anisotropic and the isotropic orientational distribution function and $F_{DY}^{ex}(\rho,T)$ is the excess dipolar Yukawa free energy for an isotropic distribution given by the last three terms in Eq. (\ref{fdyuk}). Using the MSA direct correlation function of the DY fluid (Eq. (\ref{cdy})) within this approximation the anisotropic excess free energy functional $F^{ex}_{DY}[\rho,\{\alpha(\theta)\},T]$ can be calculated analytically, which allows one to obtain $\alpha(\theta)$ by minimizing the grand canonical functional in Eq. (\ref{gra}). From the orientational distribution function the polarization $P$ (the direction of which coincides with the direction of the external electric field) can be obtained as
\begin{equation}
P=\rho\int{d\omega}\alpha(\theta)m\cos\theta.
\label{pola}
\end{equation}
Based on the orientational distribution function $\alpha(\theta)$ Eq. (\ref{pola}) leads to the following polarization function:
\begin{equation}
P=m\rho{L}\left(\beta{m}E+3P\frac{(1-q(-\xi(y)))}{m\rho}\right)
\label{pol},
\end{equation}
where $L(x)=\coth x-1/x$ is the well known Langevin function \cite{bo1}. This is an implicit equation for the external field dependence of the polarization. The details of this calculation can be found in Ref. \cite{sz1} (where the equivalent magnetic language was adopted, i.e., the particles carry magnetic dipole moments and interact with an applied magnetic field). 

\subsection{Nonlinear dielectric effect}

In order to obtain the nonlinear dielectric permittivity $\lambda$ the third 
order derivative of the polarization with respect to the Maxwell field has to be calculated (see Eq. (\ref{der3})).
To this end we introduce the following dimensionless form of Eq. (\ref{pol}):
\begin{equation}
p(e)=L\left(e+{\Gamma}p(e)\right),
\label{main}
\end{equation}
where 
\begin{equation}
p=P/(m\rho),\,\,\,\,\, e={\beta}mE
\end{equation} 
are the dimensionless polarization and electric field strength, respectively, and $\Gamma=3(1-q(-\xi(y)))$ is a field independent parameter.
From Eq. (\ref{main}) one obtains
\begin{equation}
\frac{dp}{de}=\frac{dL(x)}{dx}{\Bigg|}_{x=a}\left(1+\Gamma\frac{dp}{de}\right),
\label{d1}
\end{equation}
where $a=e+\Gamma{p(e)}$. The corresponding second order derivative is
\begin{equation}\
\frac{d^2p}{de^2}=\frac{d^2L(x)}{dx^2}{\Bigg|}_{x=a}
\left(1+\Gamma\frac{dp}{de}\right)^2+
\Gamma\frac{dL(x)}{dx}{\Bigg|}_{x=a}\frac{d^2p}{de^2}.
\label{d2}
\end{equation}
For the third order derivative one has
\begin{eqnarray}
\frac{d^3p}{de^3}=\frac{d^3L(x)}{dx^3}{\Bigg|}_{x=a}\left(1+
\Gamma\frac{dp}{de}\right)^3+\nonumber\\
\Gamma\frac{dL(x)}{dx}{\Bigg|}_{x=a}\frac{d^3p}{de^3}+\nonumber\\
3\Gamma\frac{d^2L(x)}{dx^2}{\Bigg|}_{x=a}\left(1+\Gamma\frac{dp}{de}\right)
\frac{d^2p}{de^2}.
\label{d3}
\end{eqnarray}
Since we consider thermodynamic states without spontaneous polarization (see Ref. \cite{sz1}) one has $L(x=0)=0$ so that $p(e=0)=0$. With
\begin{eqnarray}
\frac{dL(x)}{dx}{\Bigg|}_{x=0}=\frac{1}{3},\,\,\,\,\,\,\,\,\,
\frac{d^2L(x)}{dx^2}{\Bigg|}_{x=0}=0,\nonumber\\
\frac{d^3L(x)}{dx^3}{\Bigg|}_{x=0}=-\frac{2}{15}
\end{eqnarray}
one finds
\begin{equation}
\frac{dp}{de}{\Bigg|}_{e=0}=\frac{1}{3-\Gamma}=\frac{1}{3q(-\xi(y))},\
\label{d01}
\end{equation}
\begin{equation}
\frac{d^2p}{de^2}{\Bigg|}_{e=0}=0,
\label{d02}
\end{equation}
and
\begin{equation}
\frac{d^3p}{de^3}{\Bigg|}_{e=0}=-\frac{54}{5}\frac{1}{(3-\Gamma)^4}
=-\frac{2}{15}\frac{1}{q^4(-\xi(y))}\,\,.
\label{d03}
\end{equation}
Equations (\ref{ep0}), (\ref{yp}), and (\ref{d01}) render back the expression in Eq. (\ref{ep0msa}) for the linear dielectric permittivity. For the nonlinear dielectric permittivity Eqs. (\ref{der3}) and (\ref{d03}) yield
\begin{eqnarray}
\lambda=-\frac{m^4\rho}{45(k_BT)^3}\frac{1}{q^4(-\xi(y))}=\nonumber\\
-\frac{m^2}{20\pi(k_BT)^2}\frac{y}{q^4(-\xi(y))}.
\label{lammsa}
\end{eqnarray}
We note that for $\Gamma=0$ Eq. (\ref{main}) reduces to the Langevin equation for non-interacting dipoles,
\begin{equation}
p(e)=L(e),
\end{equation}
while for $\Gamma=3y$ one obtains the well known mean field Debye-Weiss polarization equation
\begin{equation}
p(e)=L(e+3yp(e)).
\end{equation}
Accordingly, for non-interacting dipoles (i.e., $\Gamma=0$ and $q(-\xi)=1$) one has for the linear and nonlinear dielectric permittivities
\begin{eqnarray}
\epsilon_0=1+3y,\nonumber\\
\lambda=-\frac{m^4\rho}{45(k_BT)^3}=
-\frac{m^2}{20\pi(k_BT)^2}y,
\label{LA}
\end{eqnarray}
respectively, and within the framework of the Debye-Weiss theory (i.e., $\Gamma=3y$ and $q(-\xi)=1-y$)
\begin{eqnarray}
\epsilon_0=1+3\frac{y}{1-y},\nonumber\\
\lambda=
-\frac{m^4\rho}{45(k_BT)^3}\frac{1}{(1-y)^4}=\nonumber\\
-\frac{m^2}{20\pi(k_BT)^2}\frac{y}{(1-y)^4}.
\label{DW}
\end{eqnarray}
The latter equations show that not only the linear but also the nonlinear dielectric permittivity diverge for $y=4\pi\rho{m^2}/(9k_BT)\rightarrow{1}$, which for a fixed   value $m$ of the dipole moment provides the Debye-Weiss critical temperature $k_BT_c=4\pi\rho{m^2}/9$ of the isotropic liquid - ferroelectric liquid second-order phase transition.
In the following these theoretical predictions will be compared with corresponding Monte Carlo simulation results.
\section{Monte Carlo simulations}
We have performed Monte Carlo simulations for DY fluids using the canonical NVT ensemble and applying Boltzmann sampling, periodic boundary conditions, and the minimum image convention \cite{al2}. 
In order to take into account the long-ranged character of the dipolar interaction the so-called reaction field (RF) method is used. According to this method we consider our system to be a macroscopic spherical sample composed of a number of replicas of the basic simulation cell embedded in a dielectric continuum with dielectric constant $\epsilon_{RF}$. In this case in the spherical sample the 
internal (Maxwell) field $\mathbf{E}$ is not equal to the external applied field 
$\mathbf{E}_0$, but \cite{ne1}
\begin{equation}
\mathbf{E}=\left(\frac{2\epsilon_{RF}+1}{2\epsilon_{RF}+\epsilon_E}\right)
\mathbf{E}_0.
\label{nau}
\end{equation}
For such a system, using the third order Taylor series expansion of the polarization with respect to the external field, Kusalik \cite{ku1} has shown that
\begin{eqnarray}
\epsilon_{E}\simeq\epsilon_0+\nonumber\\
\left(\frac{\epsilon_0+2\epsilon_{RF}}
{2\epsilon_{RF}+1}\right)^2
\left[\frac{4\pi\beta^3}{90V}\left(3\langle{\mathbf{M}^4}\rangle_0
-5\langle{\mathbf{M}^2}\rangle^2_0\right)\right]E_0^2,
\label{enl}
\end{eqnarray}
where
\begin{equation}
\epsilon_0=\frac{1+2\epsilon_{RF}+\frac{8\pi\beta\epsilon_{RF}}{3V}
\langle{\mathbf{M}^2}\rangle_0}
{1+2\epsilon_{RF}-\frac{4\pi\beta}{3V}\langle{\mathbf{M}^2}\rangle_0}.
\label{el}
\end{equation}
In Eqs. (\ref{enl}) and (\ref{el}) $\mathbf{M}$ is the total dipole moment of the system of volume $V$,
\begin{equation}
\mathbf{M}=\sum_{i=1}^N{\mathbf{m}_i},
\end{equation}
and $\langle{\mathbf{M}^2}\rangle_0$, $\langle{\mathbf{M}^4}\rangle_0$ are the ensemble averages of the corresponding moments in zero external field.

In our simulations we apply a conducting boundary condition which means $\epsilon_{RF}\rightarrow\infty$. In this limit Eqs. (\ref{nau}), (\ref{enl}), and (\ref{el}) for the internal field dependent dielectric permittivity lead to
\begin{equation}
\epsilon_E=\epsilon_0+\frac{4\pi\beta^3}{90V}
\left(3\langle{\mathbf{M}^4}\rangle_0-5\langle{\mathbf{M}^2}\rangle_0^2\right)E^2,
\label{nde}
\end{equation}
with
\begin{equation}
\epsilon_0=1+\frac{4\pi\beta}{3V}\langle{\mathbf{M}^2}\rangle_0.\
\label{de}
\end{equation}
Comparing Eq. (\ref{nde}) with Eq. (\ref{lam}) for the nonlinear dielectric permittivity we obtain
\begin{equation}
\lambda=\frac{\beta^3}{90V}\left(3\langle{\mathbf{M}^4}\rangle_0-
5\langle{\mathbf{M}^2}\rangle_0^2\right).
\label{lam2}
\end{equation}
In our NVT ensemble MC simulations $\epsilon_0$ and $\lambda$ are calculated from 
Eqs. (\ref{de}) and (\ref{lam2}), respectively.

A spherical cutoff of the hard core DY pair potentials at half of the cubic box has been applied  and long-ranged corrections (LRC) were taken into account \cite{al2}. In our simulations $N=256$ particles have been used. We have not resorted to any finite-size scaling analysis for detecting the occurrence of the isotropic - anisotropic phase transitions. The simulations were started from a hcp lattice configuration with randomly oriented dipoles. After 20.000 equilibration cycles, $2\times10^6$ - $4\times10^6$ production cycles were used. Statistical errors were calculated from the standard deviations of sub-averages containing $2\times10^5$ cycles.

\section{Results and discussion}
In the following we shall use reduced quantities: $T^*=k_BT/\varepsilon_Y$ as the reduced temperature, $\rho^*=\rho\sigma^3$ as the reduced density, 
$m^*=m/\sqrt{\varepsilon_Y\sigma^3}$ as the reduced dipole moment, and $\lambda^*=\lambda\varepsilon_Y/\sigma^3$ as the reduced nonlinear dielectric permittivity.
Concerning the range of the Yukawa potential, all our results correspond to $\kappa=1.8/\sigma$.

Figure 1 shows the linear and nonlinear dielectric permittivity as function of the reduced number density $\rho^*$. For $(m^*)^2=0.5$ and $T^*=1$ the predictions of our MSA theory (see Eqs. (\ref{ep0msa})-(\ref{yp}) and (\ref{lammsa})) 
\begin{figure}[ht]
\includegraphics*[width=17pc]{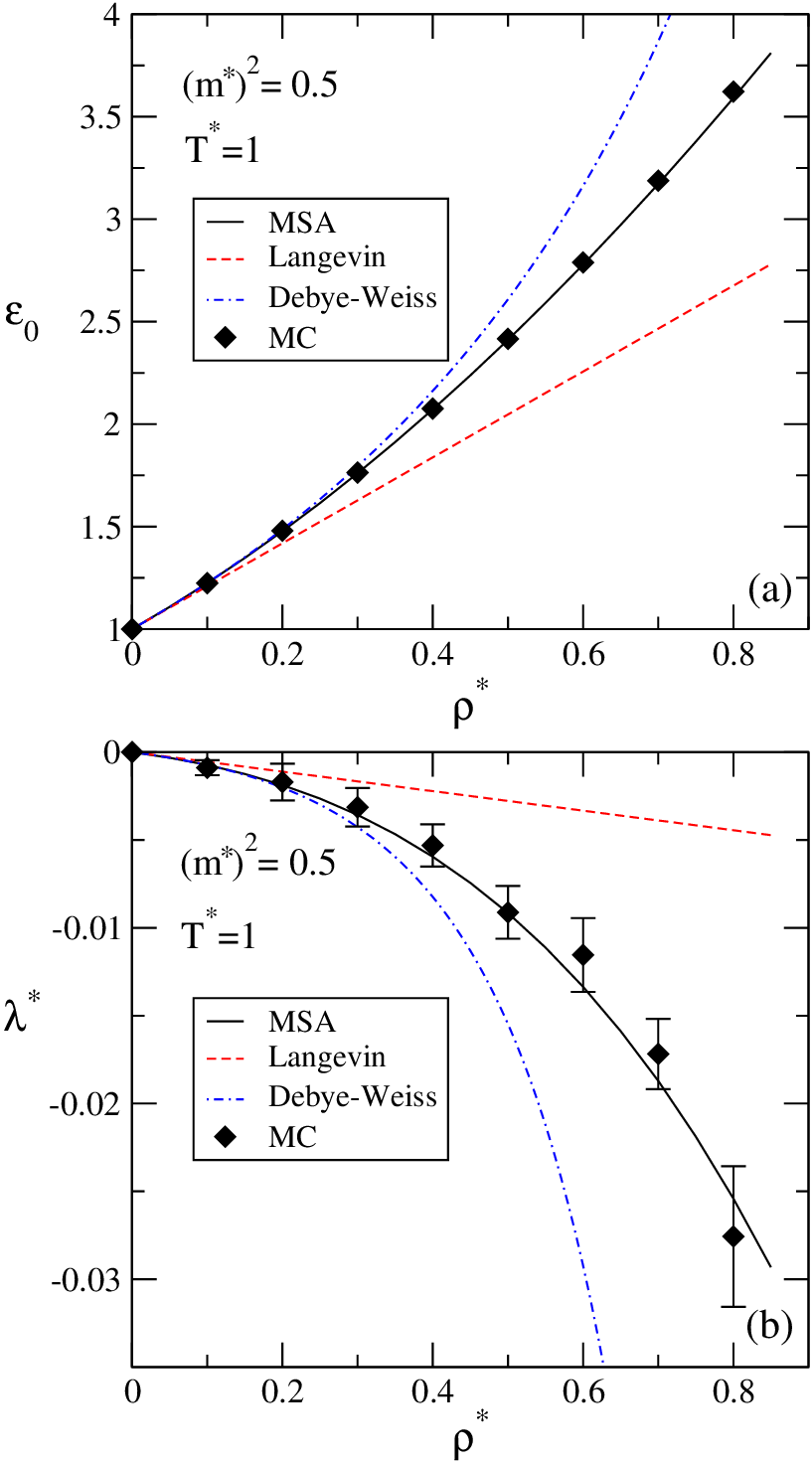}
\begin{spacing}{1.2}
\caption{Linear (a) and nonlinear (b) MSA dielectric properties of hard core dipolar Yukawa fluids for $(m^*)^2=0.5$ and $T^*=1$ in comparison with the corresponding Langevin and Debye-Weiss approximations. The symbols in both figures represent our Monte Carlo simulation data whereas the lines are obtained from the three corresponding theories. In (a) the size of the error bars is that of the symbols. Within the Debye-Weiss theory both $\epsilon_0$ and $\lambda^*$ diverge at $\rho_c^*=1.432$. Note that according to Eqs. (\ref{lamser}) and (\ref{diff2}) $\lambda^*$ vanishes linearly for $\rho^*\rightarrow{0}$.}
\end{spacing}
\end{figure}
are compared 
with the Langevin (Eq. (\ref{LA})) and Debye-Weiss (Eq. (\ref{DW})) approximations and our MC data. The Langevin theory predicts, in general, a linear dependence of $\epsilon_0$ and $\lambda$ on $\rho$. Figure 1 shows that the interparticle interaction enhances $\epsilon_0$ and $\rvert\lambda^*\rvert$ relative to the corresponding values of the Langevin theory. The Debye-Weiss theory overestimates the MSA results for $\epsilon_0$ and $\rvert\lambda^*\rvert$.
The critical density, at which within the Debye-Weiss theory $\epsilon_0$ and $\lambda$ diverge, is $\rho_c=1.432$. 
The behavior of $\epsilon_0$ and $\lambda$ shown in Fig. 1 for various approximations can be understood in terms of a low density expansion which amounts to an expansion in terms  of $y<1$ which is proportional to $\rho$ (see Eq. (\ref{yyy})). Within MSA the linear dielectric permittivity of the hard core dipolar Yukawa fluid has the expansion \cite{sz3}
\begin{equation}
(\epsilon_0)_{MSA}=1+3y+3y^2+\frac{3}{16}y^3+O(y^4).
\label{msaser}
\end{equation}
From Eqs. (\ref{ep0msa}), (\ref{yp}), and (\ref{msaser}) one finds
\begin{equation}
\frac{1}{q(-\xi(y))}=\frac{\epsilon_0-1}{3y}=1+y+\frac{1}{16}y^2+O(y^3), 
\end{equation}
so that
\begin{eqnarray}
\lambda_{MSA}=-\frac{m^2}{20\pi(k_BT)^2}\frac{y}{q^4(-\xi(y))}=\nonumber\\
-\frac{m^2}{20\pi(k_BT)^2}(y+4y^2+\frac{25}{4}y^3)+O(y^4).
\label{lamser}
\end{eqnarray}
The corresponding expansions within the Debye-Weiss approximation are
\begin{eqnarray}
(\epsilon_0)_{DW}=1+3\frac{y}{1-y}=\nonumber\\
1+3y+3y^2+3y^3+O(y^4)=\nonumber\\
(\epsilon_0)_{MSA}+\frac{45}{16}y^3 +O(y^4)
\label{diff1}
\end{eqnarray}
and
\begin{eqnarray}
\lambda_{DW}=-\frac{m^2}{20\pi(k_BT)^2}\frac{y}{(1-y)^4}=\nonumber\\
-\frac{m^2}{20\pi(k_BT)^2}(y+4y^2+10y^3)+O(y^4)=\nonumber\\
\lambda_{MSA}-\frac{3m^2}{16\pi(k_BT)^2}y^3+O(y^4).
\label{diff2}
\end{eqnarray}
Equations (\ref{diff1}) and (\ref{diff2}) explain the trends of the various curves shown in Fig. 1. For the present choice of $m^*$ and $T^*$ the MC data and the MSA results for the linear dielectric permittivity agree very well. This agreement remains rather good also for the nonlinear dielectric permittivity. We note that the very good agreement between the MSA and the MC simulation data for the linear dielectric permittivity has also been reported in Ref. \cite{sz3}.
\begin{figure}[ht]
\includegraphics*[width=17pc]{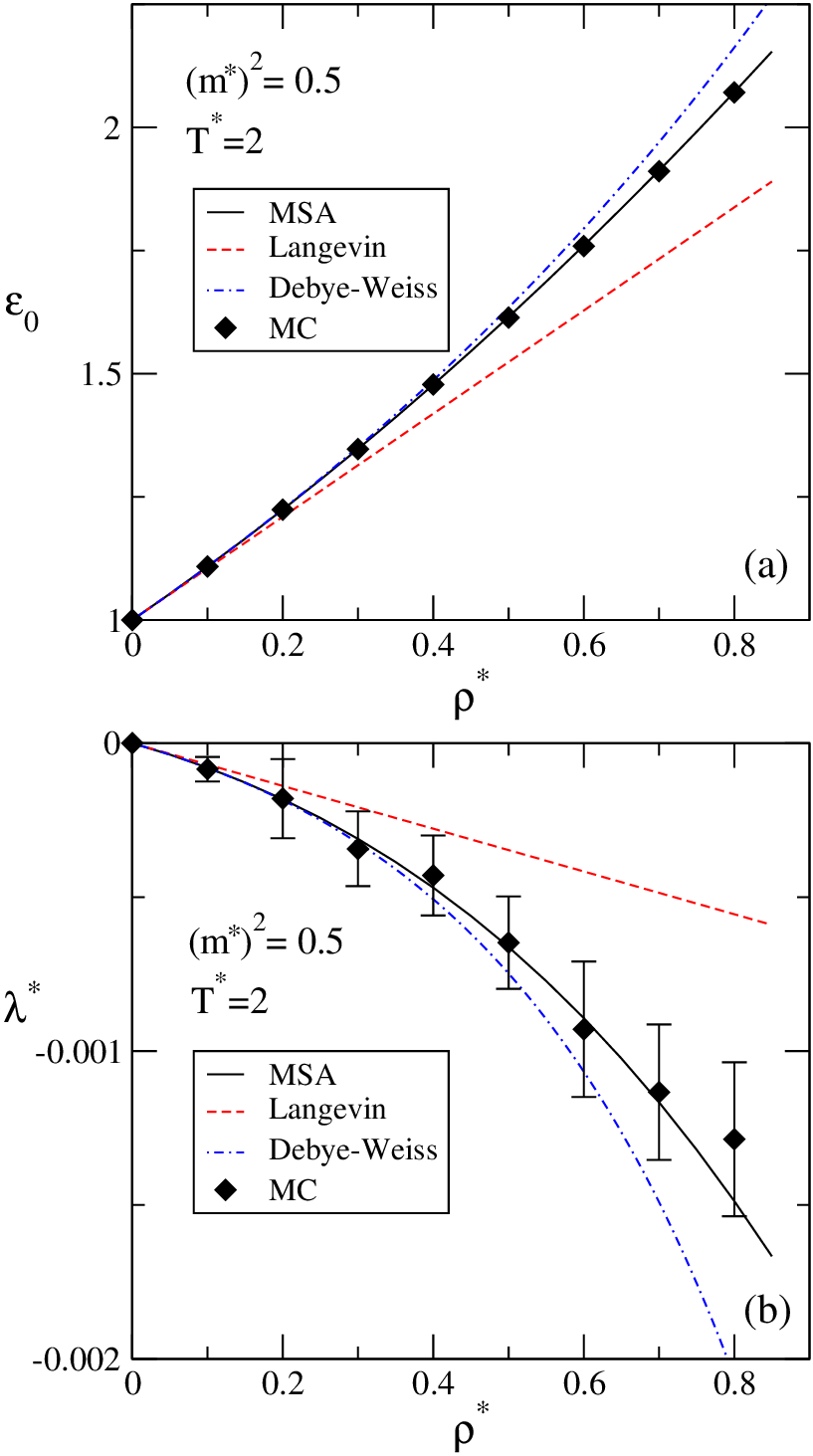}
\begin{spacing}{1.5}
\caption{Same as in Fig. 1 for $T^*=2$. Within the Debye-Weiss theory both $\epsilon_0$ and $\lambda^*$ diverge at $\rho_c^*=2.865$.}
\end{spacing}
\end{figure}
As shown in Fig. 2, the increase of the reduced temperature from $T^*=1$ to $T^*=2$ does not change the character of the curves. The MSA results are in good agreement with the simulation data. With increasing temperature the discrepancy between the Debye-Weiss theory and the simulation data is reduced because for higher temperatures the critical density, at which within the Debye-Weiss theory $\epsilon_0$ and $\lambda$ diverge, is shifted to higher densities (here $\rho^*_c=2.865$; we note that this value of the density is physically not accessible, because the corresponding packing fraction $\eta=\pi\rho^*/6$ would be larger than 1). The absolute values of $\epsilon_0$ and $\lambda$ are considerably reduced upon raising the temperature. 
\begin{figure}[ht]
\includegraphics*[width=17pc]{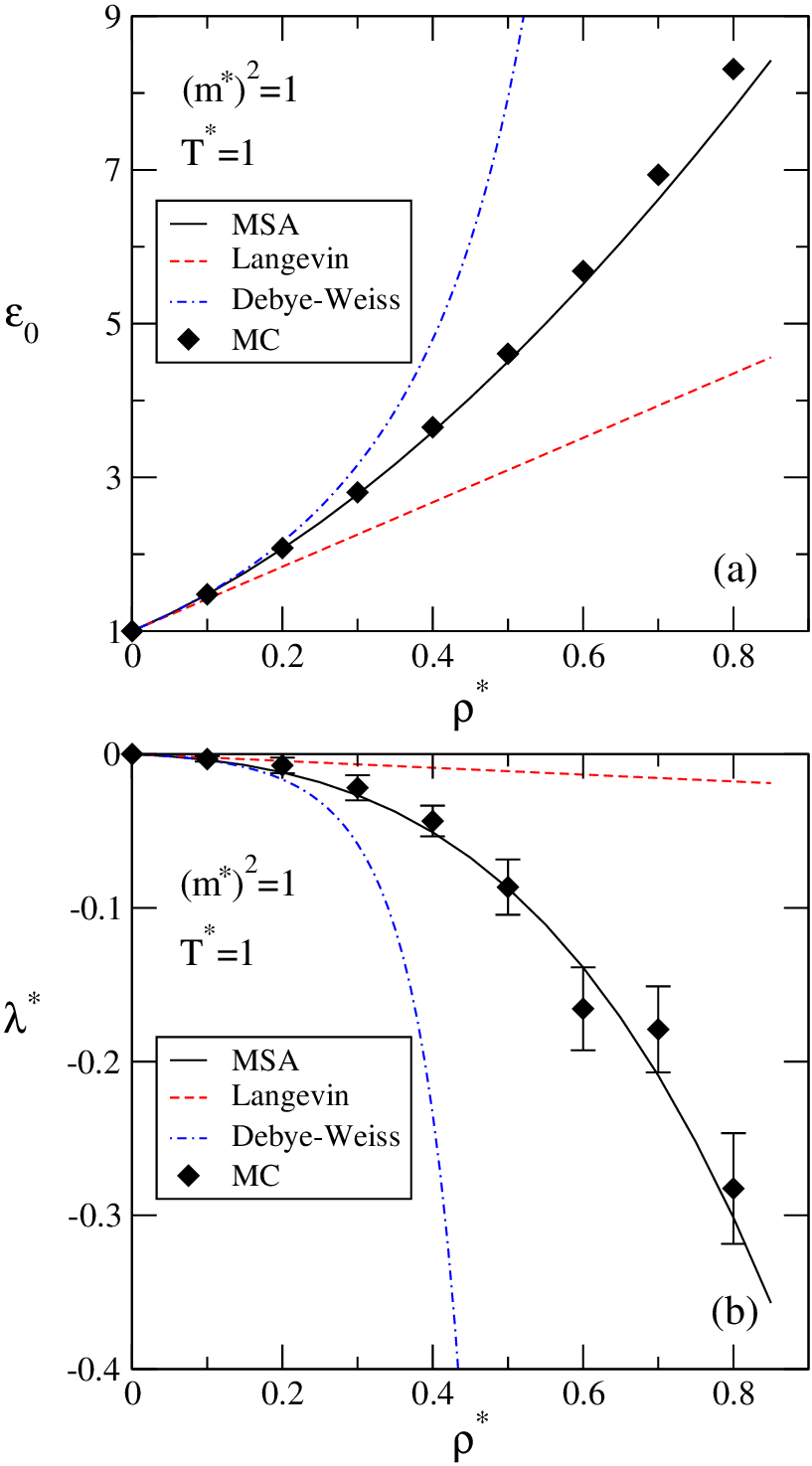}
\begin{spacing}{1.5}
\caption{Same as in Fig. 1 for $(m^*)^2=1$. Within the Debye-Weiss theory both $\epsilon_0$ and $\lambda^*$ diverge at $\rho_c^*=0.716$.}
\end{spacing}
\end{figure}
Figure 3 shows that upon increasing the dipole moment from $(m^*)^2=0.5$ to $(m^*)^2=1$ the agreement between the MSA and the Monte Carlo simulation data does not change significantly. However, the discrepancy between the Debye-Weiss theory and the simulation data has widened because due to  the increase of the dipole moment the critical density, at which within the Debye-Weiss theory $\epsilon_0$ and $\lambda$ diverge, is shifted to a lower value ($\rho^*_{c}=0.716$). The Langevin theory underestimates both $\epsilon_0$ and $\rvert\lambda^*\rvert$. The increase of the dipole moment significantly increases $\epsilon_0$ and $\rvert\lambda^*\rvert$.

Figure 4 shows that upon increasing the temperature from $T^*=1$ to $T^*=2$ at a fixed value $(m^*)^2=1$ of the dipole moment the agreement between the MSA and the simulation data is improved and the values of $\epsilon_0$ and $\rvert\lambda^*\rvert$ are significantly reduced. The discrepancy between the Debye-Weiss theory and the simulation data is decreased because the critical density $\rho^*_{c}=1.432$, at which $\epsilon_0$ and $\lambda^*$ diverge according to the Debye-Weiss theory, is increased.
\begin{figure}[ht]
\includegraphics*[width=17pc]{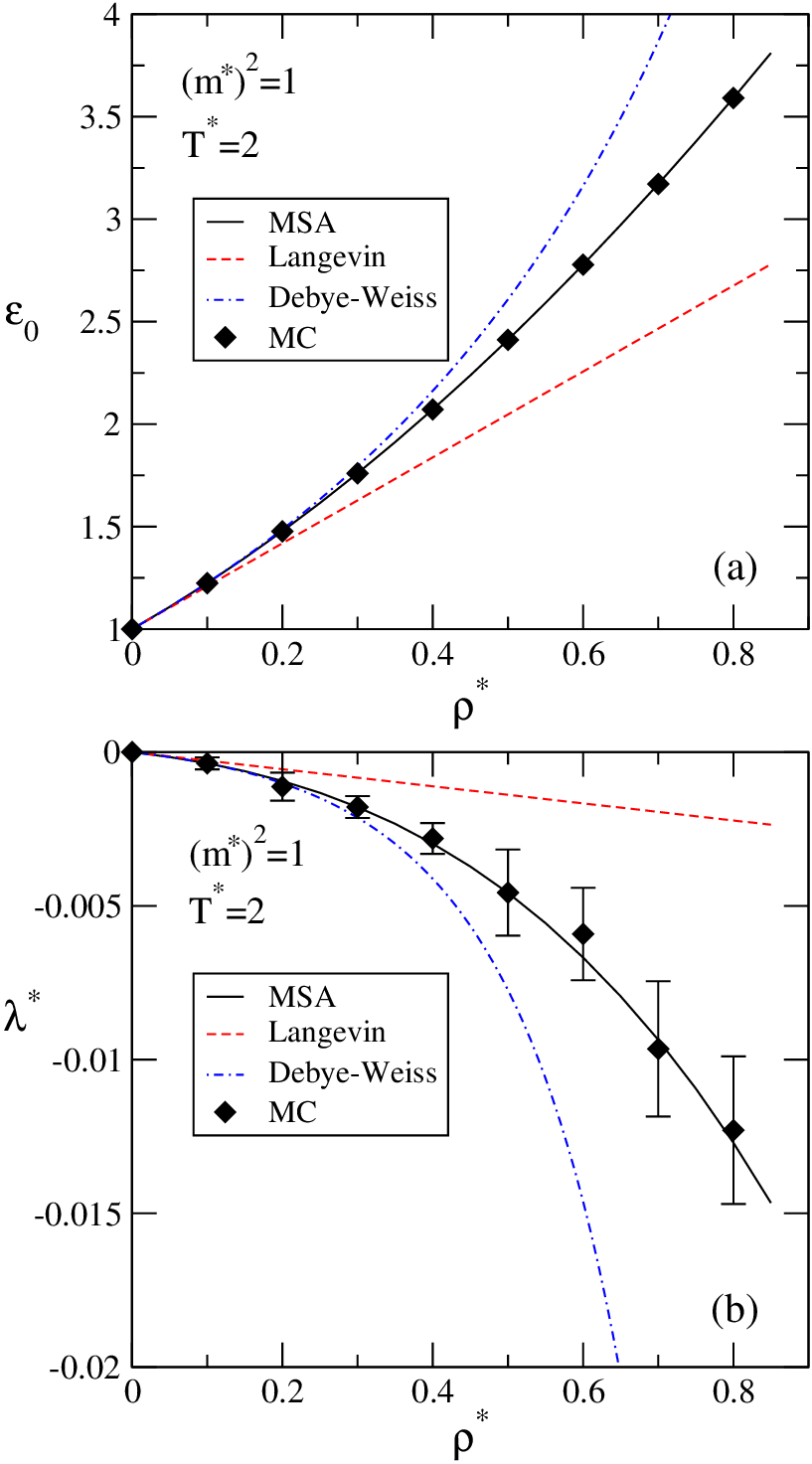}
\begin{spacing}{1.5}
\caption{Same as in Fig. 3 for $T^*=2$. Within the Debye-Weiss theory both $\epsilon_0$ and $\lambda^*$ diverge at $\rho_c^*=1.432$.}
\end{spacing}
\end{figure}
From Figs. 1-4 we conclude that for hard core dipolar Yukawa fluids the MSA describes the linear and nonlinear dielectric permittivities with an adequate accuracy up to liquid number densities $\rho^*{\lesssim}\,0.8$ (note that freezing occurs at $\rho^*\simeq{0.8}$; see Ref. \cite{kl1} for the related Stockmayer system), while the Langevin and the Debye-Weiss theory can be used only for $\rho^*{\lesssim}\,0.1$ and $\rho^*{\lesssim}\,0.2$, respectively.

Within the framework of our MSA theory and in agreement with our MC simulation data we have obtained a negative nonlinear dielectric permittivity for hard core dipolar Yukawa fluids. We note that using a cluster expansion and the quadratic hypernetted chain (QHNC) approximation Martina and Stell \cite{st1} obtained positive nonlinear dielectric permittivities for dipolar hard sphere fluids. They tried to explain the positive sign in terms of electrostriction (i.e., the dependence of $\rho$ on $E$) which is included in their theory. However, from an experimental point of view this explanation is unlikely because the effect of electrostriction is so weak \cite{kr1} that it cannot overcompensate the contribution to $\Delta\epsilon$ of  molecular orientations in the presence of an applied external field (see Ref. \cite{kr1}). Note that Eq. (\ref{lammsa}) as well as Eq. (\ref{ep0msa}) are also valid for dipolar hard spheres, which implies that within MSA the dispersion forces do not influence the dielectric properties.


\section{Summary}
In the present study of the nonlinear dielectric effect for dipolar fluids the following main results have been obtained:\\
1) We have applied an extension of the mean spherical approximation (MSA) theory to determine the internal electric field dependence of the polarization for hard core dipolar Yukawa fluids.\\
2) From the electric field dependence of the polarization analytical equations have been obtained for the linear and the nonlinear dielectric permittivity of these dipolar fluids.
The predicted nonlinear dielectric permittivity is negative, which corresponds to the so-called normal saturation effect of dielectric media.\\
3) Canonical Monte Carlo simulations have been carried out for the determination of the linear and nonlinear dielectric permittivity of dipolar Yukawa fluids. There is good agreement between the results from the MSA and the Monte Carlo simulation data for the reduced dipole moments $(m^*)^2\leq1$ studied here (see Figs. 1-4).\\
4) We have compared our theoretical results with the corresponding Langevin and Debye-Weiss approximations. The observed trends are in agreement with the low density behavior of the various approximations (see Eqs. (\ref{diff1}) and (\ref{diff2})).

Our new theoretical approach provides a quantitatively reliable description for the nonlinear dielectric permittivity. This raises the expectation that actual experimental systems can be analyzed and understood along these lines. It will be also interesting to detect model systems which exhibit anomalous dielectric saturation, i.e., a positive nonlinear dielectric permittivity.


\begin{acknowledgments} 
I. Szalai would like to thank the Hungarian Scientific Research Fund 
(Grant No. OTKA K61314) for financial support. 
\end{acknowledgments}

\end{document}